\documentstyle[aps,prc,psfig,manuscript]{revtex}

\begin{document}
\bibliographystyle{h-physrev}

\title{K-Factors in Parton Cascades at RHIC and SPS}

\author{S.A. Bass and B. M\"uller}
\address{
        Department of Physics\\
        Duke University\\
        Durham, NC, 27708-0305, USA
        }

\date{\today}

\preprint{DUKE-TH-99-192}

\maketitle

\begin{abstract}
Different treatments for the inclusion of higher-order perturbative
QCD corrections in parton based transport models of relativistic 
heavy-ion collisions are studied and their influence on experimental 
observables is investigated. 
At RHIC, particle multiplicities may vary by more than 30\%, depending
on the correction scheme. A detailed analysis of the squared relative
momentum transfers $Q^2$ to be expected in parton (re)scatterings at 
RHIC casts doubt on the applicability of pQCD based transport approaches
at the SPS and rules out the application of higher-order correction 
schemes based on a rescaling of $Q^2$ in the running coupling constant 
$\alpha_s$ at all energies.
\end{abstract}

\pagebreak

One of the primary motivations  for the study of ultra-relativistic 
heavy ion collisions is the expectation that a new form of matter 
may be created in such reactions. That form of matter, called the 
quark-gluon plasma (QGP \cite{reviews}), is the analogue within 
nuclear matter of the plasma phase of ordinary atomic matter. 
It is expected that experiments at the relativistic heavy-ion 
collider (RHIC), which is scheduled to begin operations later
this summer, will observe and identify this new state of matter
and engage in systematic studies of the properties of strongly
interacting matter at high temperatures.

Transport theory has been among the most successful approaches
to the theoretical investigation of relativistic heavy-ion
collisions. Microscopic transport models aim at describing the 
full time evolution from the initial state of the heavy-ion 
reaction, i.e. two colliding nuclei in their ground states, up 
to the freeze-out of all particles after the reaction. If this
description is to contain a quark-gluon phase, as it will be
necessary for the RHIC energy domain, both partonic and hadronic 
degrees of freedom must be included. In the partonic domain, the 
Parton Cascade Model (PCM) \cite{geiger92a,geiger95a} and other 
approaches \cite{wang91a,wang92a} based on perturbative quantum 
chromodynamics (pQCD) and the Glauber approximation to multiple 
scattering theory have been widely used. The numerical 
implementations of these models \cite{wang94a,geiger97a} make
use of the hard parton scattering cross sections and radiative 
emissions as treated in the established {\sc Pythia} event 
generator \cite{pythia}.

It is well known that higher-order corrections to leading-order 
pQCD cross sections are large and process dependent \cite{kfactor1}. 
Often these corrections are applied in the form of a so-called 
{\em K-factor} which renormalizes the leading-order pQCD cross 
section and accounts for higher-order corrections \cite{kfactor2}. 
Although K-factors have been calculated for many pQCD processes,
in many cases they have been determined phenomenologically by
comparisons with data. The choice of the K-factor is particularly
important in the case of parton cascade models 
\cite{geiger97a,bass99vni}, because it enters into the transport 
cross section and thus strongly influences the number of parton 
rescatterings. A change in the value of the K-factor does not just
lead to a multiplicative renormalization of the cross section for
an observable.

Two well-established methods have been employed to implement
the K-factor in pQCD calculations:
\begin{enumerate}
\item 
in the form of a constant multiplicative factor:
\begin{equation}
 \frac{d\sigma^{full}_{pQCD}}{dt}(Q^2) \,=\, K \times 
        \frac{d\sigma^{LO}_{pQCD}}{dt}(Q^2)
\end{equation}
with values for $K$ in the order of $1.5 \le K \le 3$ for hard 
parton-parton scattering. In this implementation the K-factor is 
independent of the virtuality scale $Q^2$ of the scattering. 
\item 
by changing the $Q^2$ scale in the running coupling constant 
$\alpha_s$ in the form
\begin{equation}
\frac{d\sigma^{full}_{pQCD}}{dt}(Q^2) \,=\, 
\frac{d\sigma^{LO}_{pQCD}}{dt}(\alpha_s(\eta \times Q^2))
\end{equation}
with values for $\eta$ in the order of $0.03 \le \eta \le 0.08$. 
\end{enumerate} 

The second method, originally proposed by Ellis et al. \cite{ellis86a}, 
introduces a strong $Q^2-$dependence, due to the functional form of 
the running coupling constant $\alpha_s$. In leading order:
\begin{equation}
\frac{d\sigma_{pQCD}}{dt} \sim \alpha^2_s \quad \mbox{and} \quad
\alpha_s^{LO} \,=\, \frac{12 \, \pi}{(33 - 2 n_f)\, \ln(Q^2/\Lambda^2)}
\end{equation} 

\begin{figure}
\centerline{\psfig{figure=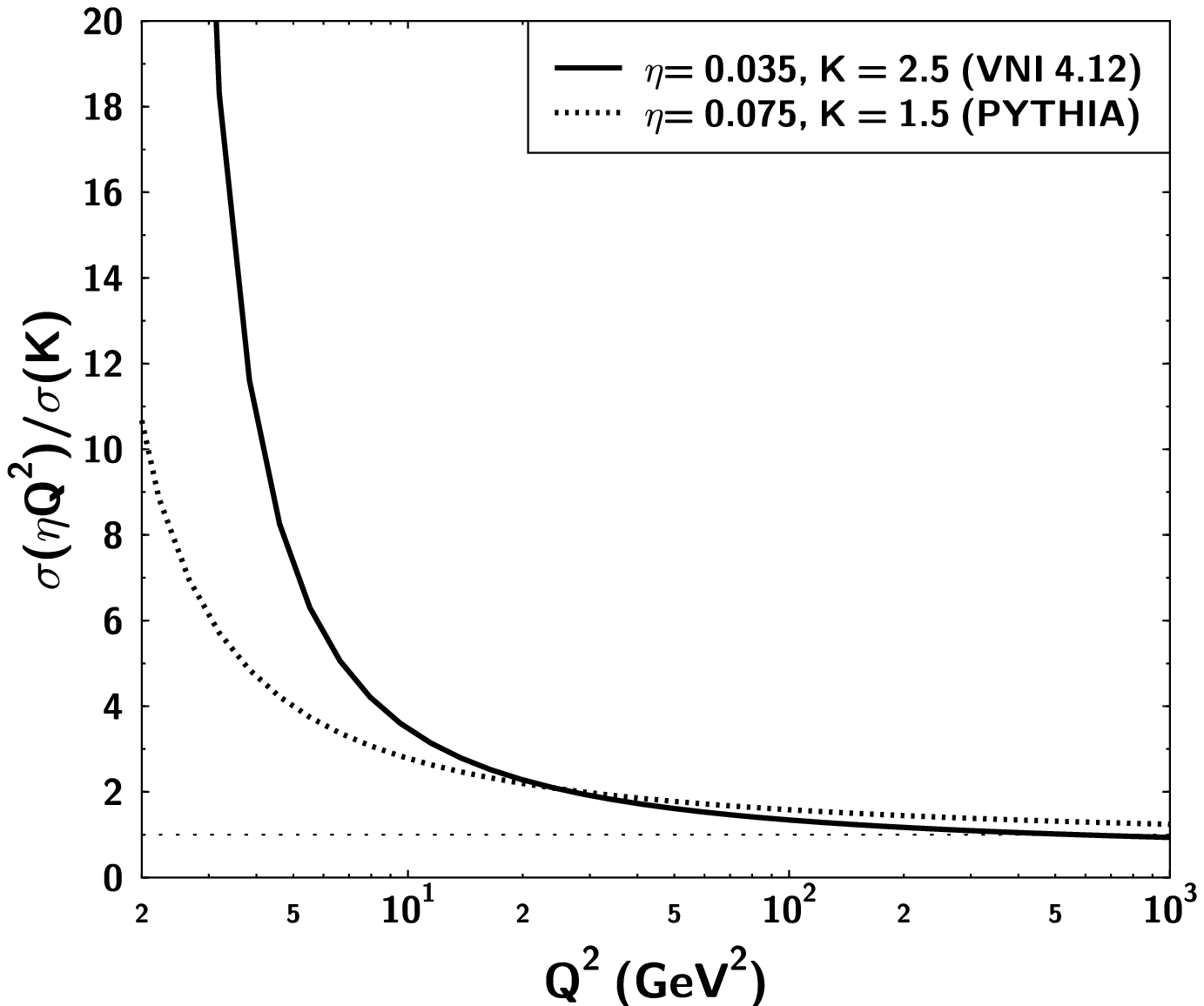,width=5in}}
\caption{\label{sigma_rat}
Ratio of the hard parton-parton scattering cross section
$\sigma(\eta Q^2)$ with higher-order corrections expressed as
a scale change in $\alpha_s(Q^2)$ (solid line), and the cross 
section corrected with a constant K-factor, $\sigma(K)$ (dotted
line) versus the squared momentum-transfer $Q^2$. The $Q^2$ scale 
renormalization leads to a divergent cross section at small,
but still reasonable values of $Q^2$.}
\end{figure}

How do the two methods compare to each other? Figure~\ref{sigma_rat}
shows the ratio of the hard parton-parton scattering cross section
corrected by a scale change in $Q^2$, $\sigma(\eta Q^2)$ and the 
scattering cross section corrected by a constant {\em K-factor}, $\sigma(K)$,
as a function of the squared momentum transfer $Q^2$. The solid
line depicts the ratio with values for $\eta$ and $K$ used in the 
parton cascade model (code version VNI~4.12) \cite{geiger97a,bass99vni}. 
The dotted line shows the same ratio with parameter values as suggested 
in {\sc Pythia} (code version 5.7)\footnote{The $Q^2$-rescaling 
method is the default in VNI, while the default in {\sc Pythia} is 
$K=1$ and higher-order corrections are an optional choice.}.
For high values of $Q^2$, on the order of 100 GeV$^2$, the cross section
ratio converges to unity and both correction methods yield the same 
result. For small $Q^2$ values ($Q^2 \le 10$ GeV$^2$), however, the 
$Q^2$ rescaling method leads to a rapidly increasing cross section 
that diverges prematurely. 
Therefore, one should consider the lower boundary for
the applicability of the $Q^2$-rescaling method to be at approximately 
$Q^2 \approx 100$~GeV$^2$, at which the deviation between the two correction
methods exceeds 50\%.

\begin{figure}
\centerline{\psfig{figure=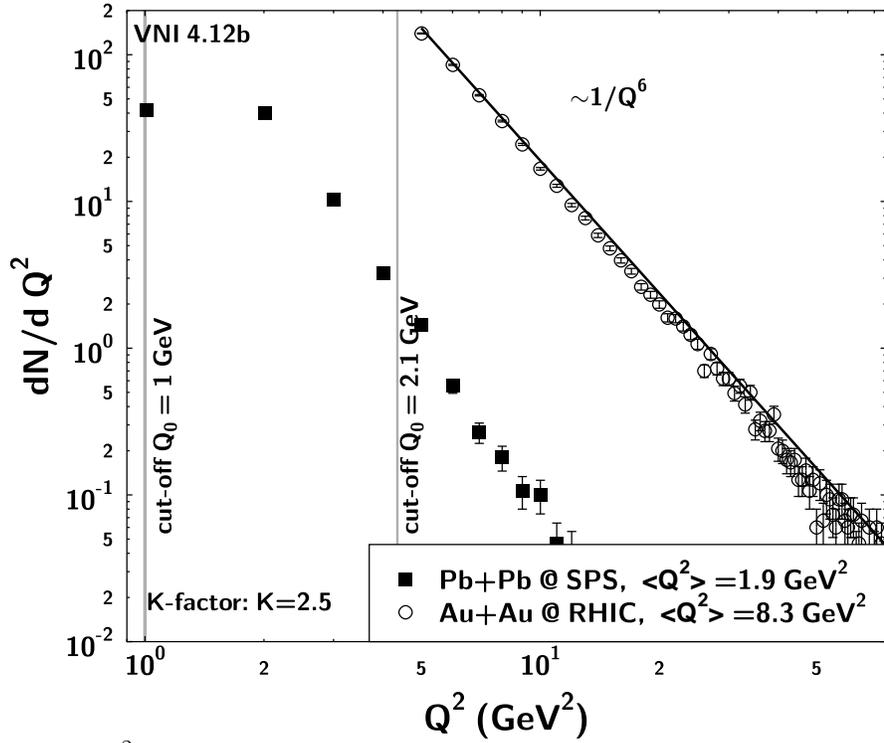,width=5in}}
\caption{\label{dndq2}
$dN/d(Q^2)$ distribution of hard parton-parton scatterings 
in central Au+Au collisions at RHIC (open circles) and in central 
Pb+Pb collisions at the CERN/SPS (full squares). The calculations
were performed with a constant K-factor ($K=2.5$). 
The distribution shows an asymptotic behavior 
$d\sigma/dQ^2 \sim 1/Q^6$ for large $Q^2$, the distribution at 
the SPS is strongly cut-off dominated.}
\end{figure}

The choice of the correction scheme is always critical in transport
theoretical applications, because a wide range of $Q^2$ values is probed. 
Figure~\ref{dndq2} shows the $Q^2$ distribution $dN/d(Q^2)$ of ``hard'' 
parton-parton scatterings calculated for central Au+Au collisions at 
RHIC (open circles). The $Q^2$ distribution for central Pb+Pb collisions 
at the Super-Proton-Synchrotron (SPS) at CERN is also shown (black squares). 
Both calculations were made with the parton cascade model code VNI~4.12, 
using the constant K-factor option with $K=2.5$. 
The $Q^2$-scale is defined as 
$Q^2 =\hat{s}$ for $s$-channel processes and as  
$Q^2 = \frac{1}{2} (m^2_{\perp 1}+m^2_{\perp 2})$ for $t$- and
$u$-channel processes.
At RHIC, the average 
squared momentum transfer $\langle Q^2 \rangle$=8.3~GeV$^2$ 
and the distribution exhibits a power-law behavior $d\sigma/dQ^2\sim 1/Q^6$,
which is due to the Rutherford-like ($d\sigma/dQ^2\sim 1/Q^4$) form of the
leading-order pQCD cross section and the parton distribution functions which
add additional inverse powers of $Q$.

In the limit of small $Q^2$ values the distribution is dominated by the 
cut-off value $Q_0 = 2.1$~GeV. For a constant K-factor the average value 
of the strong coupling constant $\alpha_s(Q^2)$ probed in these
hard parton-parton collisions is $\langle \alpha_s \rangle = 0.29$. 
If the $Q^2$ rescaling method is used instead, the
number of hard parton-parton scatterings increases strongly, 
since the hard parton-parton cross section increases by a factor
of four at $\langle Q^2 \rangle \approx 8$~GeV$^2$ (see 
Fig.~\ref{sigma_rat}). The average value of the strong coupling 
constant probed with the rescaled values of $Q^2$ is 
$\langle \alpha_s \rangle = 0.96$. It is therefore highly questionable
whether the $Q^2$ rescaling method for the higher-order
pQCD corrections is suitable at RHIC energies.

The situation at the SPS is even more dramatic: the average squared momentum 
transfer $\langle Q^2 \rangle$ here is only 1.9~GeV$^2$. 
Again, the distribution is limited for small $Q^2$ by the cut-off parameter
which is now chosen to be $Q_0 = 1.0$~GeV. 
Obviously, the entire
distribution in the SPS energy regime is dominated by the choice of the 
cut-off parameter $Q_0$ and therefore the applicability of pQCD in general
and the PCM ansatz in particular is very questionable. The average value 
of the strong coupling constant $\alpha_s$ probed here is 
$\langle \alpha_s \rangle = 0.39$. Employing the $Q^2$ rescaling method 
at $\langle Q^2 \rangle = 1.9$~GeV$^2$ leads to a hard parton-parton
cross section two orders of magnitude larger than for the choice $K=2.5$,
and $\langle \alpha_s \rangle$ probed with the rescaled values of $Q^2$ 
increases to 9.3, clearly well beyond the domain of applicability of pQCD.

\begin{figure}
\centerline{\psfig{figure=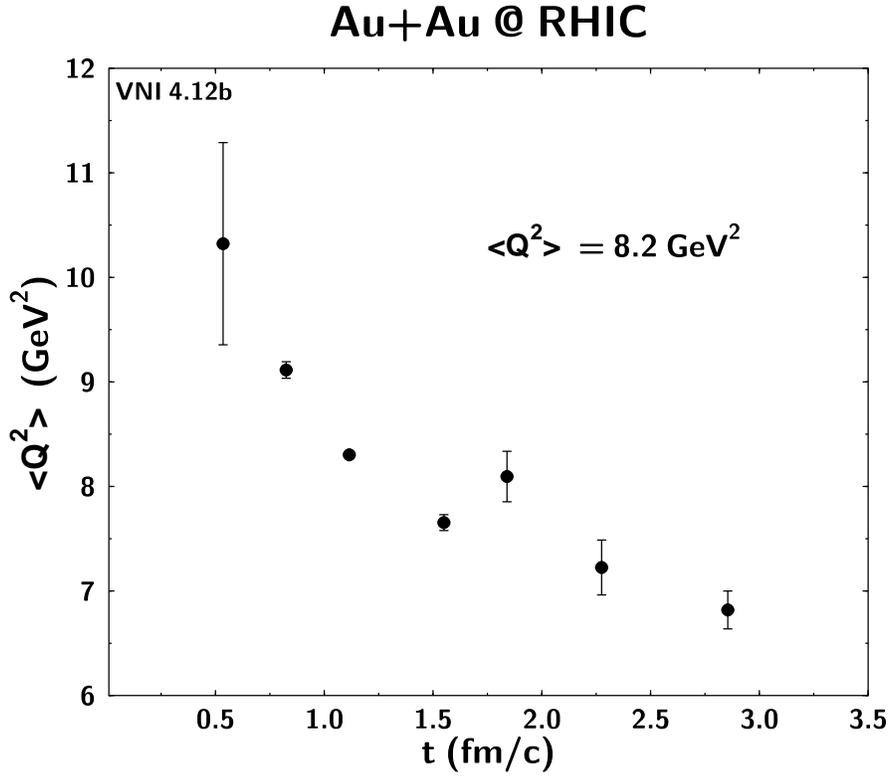,width=5in}}
\caption{\label{q2_t}
Time-evolution of the average squared momentum transfer 
$\langle Q^2 \rangle$ in hard parton-parton interactions 
for central Au+Au collisions at RHIC. The strong decrease 
of $\langle Q^2 \rangle$ and subsequent saturation are an 
indication of the onset of local thermalization.  The strong 
time-dependence does not permit the choice of an unambiguous 
fixed $Q^2$ scale for the entire reaction.}
\end{figure}

The situation with respect to the distribution of $Q^2$ values is even
more complex if one takes the initial non-equilibrium nature of an
ultra-relativistic heavy-ion collision into account: initial, 
parton-parton scatterings are bound to occur at far larger values of
$Q^2$ than rescatterings in the later reaction stages 
when a thermalized parton gas
has been formed. Figure~\ref{q2_t} shows the time-evolution of
the average squared momentum transfer $\langle Q^2 \rangle$ for central
Au+Au collisions at RHIC. The analysis is based on a VNI~4.12 calculation,
again utilizing a constant K-factor of $K=2.5$. Early rescatterings
of incident partons occur at rather large values of $Q^2 \ge 10$~GeV$^2$.
However, between the time $t_{\rm c.m.} \approx 0.5$~fm/c and 
$t_{\rm c.m.} \approx 
1.5$~fm/c the average $\langle Q^2 \rangle$ decreases rapidly and then 
saturates at a value of about $7-8$~GeV$^2$ until hadronization sets in.
The time span until saturation, $\Delta t_{\rm c.m.} \approx 1$~fm/c, is in
agreement with previous predictions concerning the thermalization time
of a deconfined parton gas formed in ultra-relativistic heavy-ion 
collisions \cite{eskola94a}. Most importantly, however, the strong
time-dependence of $\langle Q^2 \rangle$ shows that there is no single 
$Q^2$ scale which can be chosen to unambiguously set the pQCD scale 
in an ultra-relativistic heavy-ion collision.

\begin{figure}
\centerline{\psfig{figure=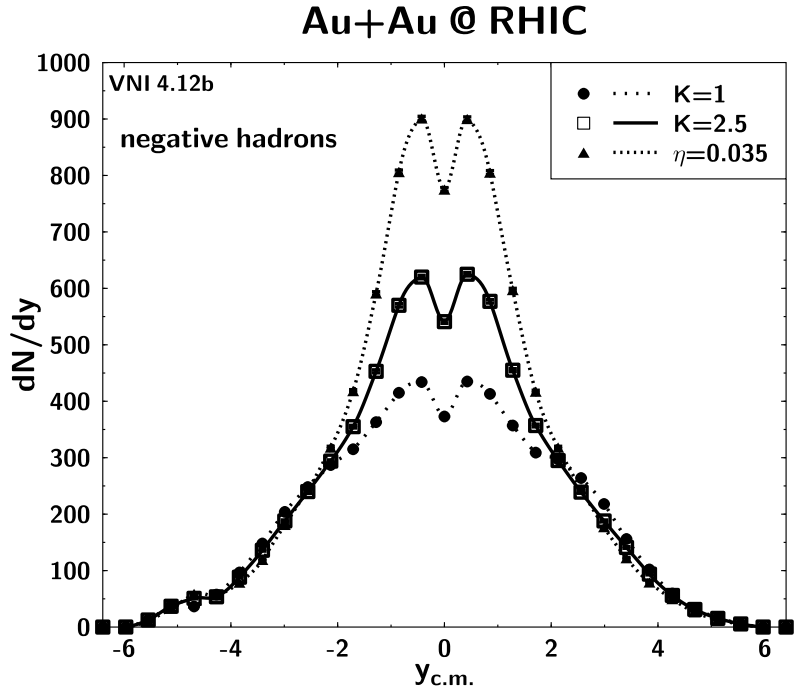,width=4.5in}}
\caption{\label{dndy}
Rapidity distribution of negatively charged hadrons ($h^-$)
from VNI~4.12, calculated with K-factors of $K=1.0$ (full circles), 
$K=2.5$ (open squares), and $Q^2$ rescaling (full triangles).}
\end{figure}

How does the choice of the higher order correction scheme influence
experimentally observable quantities at RHIC? Figure~\ref{dndy} shows
the rapidity distribution of negatively charged hadrons, calculated
with VNI~4.12 and a constant K-factor $K=1.0$ (full circles), $K=2.5$
(open squares) and $Q^2$ rescaling (full triangles). Introducing 
higher-order pQCD corrections via a constant K-factor $K=2.5$ leads
to an increase in the $h^-$ rapidity-density at $y_{c.m.}$ 
of roughly 50\%, performing a
correction based on a $Q^2$ rescaling would increase the multiplicity
by an additional 30\%. Notably, the changes only affect the area around
mid-rapidity, $-2\le y_{\rm cm}\le 2$ since the beam- and target-rapidity
regions are dominated by soft beam-jet physics and are not affected
by the amount of hard parton rescattering. 
Even though the difference in
the hard parton-parton cross section between the two methods is a factor
of four at $\langle Q^2 \rangle \approx 8$~GeV$^2$, the change in final
hadron multiplicity amounts to only 30\%. This ``reduction'' in the observable
effect compared to the change in cross section is caused by the interplay of
hard scatterings and soft beam/target-jet fragmentation processes. An increase
in the amount of hard scattering may lead to a decrease in the soft contribution
so that the sum of both is lower than one may expect from the scaling of
the hard cross section alone.

Based on the above findings, the use of the $ Q^2$ rescaling method 
for implementing higher-order pQCD corrections should be regarded as 
unacceptable in the framework of the parton cascade model. A major 
revision of the PCM code VNI, which has employed this method as the
default, is in progress \cite{vni_bms}. A forthcoming publication will 
present a thorough analysis of elementary hadron-hadron data for the 
determination of the appropriate phenomenological K-factor, as well 
as a full set of predictions for nucleus-nucleus collisions at RHIC. 
One of the many intriguing issues to be addressed is the question 
whether the hot and dense partonic matter state created in these
collisions introduces a medium-dependence into the K-factor, or 
-- more generally speaking -- whether it will be possible to probe 
the medium properties of hot and dense QCD-matter at RHIC by means
of pQCD processes. 

In summary, we have investigated  
different higher-order pQCD correction schemes currently
used in transport models in the context
of ultra-relativistic heavy-ion collisions at RHIC and SPS.  
Measurable quantities like the multiplicity of negative hadrons 
at RHIC are influenced on the order of 30\%, depending on the choice 
of the correction scheme.  An analysis of the $Q^2$ regime to be 
expected at RHIC shows that correction schemes based on a rescaling 
of $Q^2$ in the running coupling constant $\alpha_s$ yield 
unacceptable results due to the divergence of this correction 
scheme for small values of $Q^2$.  
The application of the  parton cascade model 
to nuclear collisions at the CERN SPS which were
based on numerical results obtained with the VNI implementation
of the PCM \cite{GeiSri} therefore 
need to be reconsidered.

We acknowledge helpful discussions with D.K. Srivastava, K. Eskola, 
X.N. Wang and A. Dumitru.  This work has been supported in part by 
the Alexander von Humboldt Foundation, by the U.S. Department of
Energy (grant DE-FG02-96ER40945), and by the Intel Corporation.


\pagebreak

\pagebreak

\pagebreak

\end{document}